\documentclass[11pt, oneside]{elsarticle}
\pdfoutput=1
\usepackage{lipsum}
\makeatletter
\def\ps@pprintTitle{%
 \let\@oddhead\@empty
 \let\@evenhead\@empty
 \def\@oddfoot{}%
 \let\@evenfoot\@oddfoot}
\makeatother
\RequirePackage[english=usenglishmax]{hyphsubst}
\usepackage{enumerate}
\usepackage{graphics}
\usepackage{url}
\usepackage{hyperref}	
\usepackage{breakurl}

\usepackage{float}
\usepackage{geometry}                		
\usepackage{amsmath}
\usepackage{ amssymb }
\usepackage{mathtools}
\usepackage{color}
\usepackage[table]{xcolor}
\usepackage{color,soul}
\usepackage{siunitx}
\usepackage{longtable}

\usepackage{booktabs}
\usepackage[flushleft]{threeparttable}

\usepackage{tikz,caption}
\newsavebox{\mytable}
\usetikzlibrary{fit,shapes.misc}

\usepackage{tabularx}

\usepackage{lmodern,multirow} 
\usepackage{rotating}

\newcolumntype{K}[1]{>{\centering\arraybackslash}p{#1}}

\usepackage{graphicx}				

\begin{document}

\title{Panel dataset description for econometric analysis of the ISP-OTT relationship in the years 2008-2013}
\author[label1]{Chiara Perillo\corref{cor1}}
\ead{chiara.perillo@bf.uzh.ch}
\author[label2]{Angelos Antonopoulos}
\ead{aantonopoulos@cttc.es}
\author[label2]{Christos Verikoukis}
\ead{cveri@cttc.es}

\cortext[cor1]{Corresponding author}

\address[label1]{University of Zurich, Department of Banking and Finance, Zurich, Switzerland}
\address[label2]{Telecommunications Technological Centre of Catalonia (CTTC), Castelldefels, Barcelona, Spain}

\begin{abstract}

The latest technological advancements in the telecommunications domain (e.g., widespread adoption of mobile devices, introduction of 5G wireless communications, etc.) have brought new stakeholders into the spotlight. More specifically, Over-the-Top (OTT) providers have recently appeared, offering their services over the existing deployed telecommunication networks. The entry of the new players has changed the dynamics in the domain, as it creates conflicting situations with the Internet Service Providers (ISPs), who traditionally dominate the area, motivating the necessity for novel analytical studies for this relationship. However, despite the importance of accessing real observational data, there is no database with the aggregate information that can serve as a solid base for this research. To that end, this document provides a detailed summary report for financial and statistic data for the period 2008-2013 that can be exploited for realistic econometric models that will provide useful insights on this topic. The document summarizes data from various sources with regard to the ISP revenues and Capital Expenditures (CAPEX), the OTT revenues, the Internet penetration and the Gross Domestic Product (GDP), taking into account three big OTT providers (i.e., Facebook, Skype, WhatsApp) and ten major ISPs that operate in seven different countries.

\end{abstract}


\maketitle
\vspace{0.3 cm}

\section{Introduction}

The introduction of next generation wireless communications (i.e., 4G and 5G) along with the vast proliferation of handheld smart mobile devices have motivated the appearance of Over-the-Top (OTT) providers that offer their services over the existing telecommunications networks, operated mainly by the Internet Service Providers (ISPs). The conflicted interests among these entities (e.g., same customer base, similar services, use of the same network infrastructure, etc.) have triggered a series of discussions and interactions, aiming to clarify the boundaries of the formed relationships and the obligations of each party.

The aforementioned discussions have constituted the core of the network neutrality debate, which focuses on the Internet neutrality, i.e., the equal and fair treatment of all data, without any deliberate prioritization. Although there have been some important theoretical studies to analyze the relationship between OTT providers and ISPs \cite{ma, coucheney, saucez, courcoubetis}, empirical econometric researches could provide additional intriguing insights on the debate. However, econometric studies for this particular interaction were not possible until recently, as the main explosion of OTT services took place less than ten years ago and, hence, no data were available. Moreover, obtaining real data (regarding revenues, investments and costs) is often quite complicated due to privacy concerns of the involved companies.

In the light of the above context, this report provides a detailed summary of empirical data for several important variables (in the reference period 2008-2013) that affect the relationship between OTT companies and ISPs. More specifically, we have tried to collect data with regard to the i) ISP revenues, ii) Capital Expenditures (CAPEX) for the network investments of the ISPs, iii) OTT revenues, iv) Internet penetration, and v) real Gross Domestic Product (GDP) of different countries. Our data constitute a data panel and concern ten major ISPs and three huge OTT providers that operate in seven Organization for Economic Co-operation and Development (OECD) countries (Japan, USA, UK, France, Italy, Spain and Germany). It is worth noting that, for our study, we have referred to various sources and our main goal is to provide a compact document that summarizes data that can be exploited for empirical econometric studies. Although our data are, in most cases, accurate, in case of missing data we proceeded in some reasonable estimations through interpolation (taking into account the existing data) and a comparison with the available data gives errors lower than 20\% (in particular, lower than 10\% in the 95.45\% of the cases and lower than 5\% in the 77.27\% of the cases). Moreover, in cases where specific data per country were not available (especially in case of OTT companies that operate worldwide), we resorted to some assumptions that provide a reasonable level of breakdown.

The remainder of this document is organized as follows. Section 2 provides the data regarding the ISP revenues and the CAPEX for the network investments. Section 3 provides the data for the OTT revenues. The Internet penetration and the GDP data are provided in Section 4 and 5, respectively. Finally, Section 6 concludes this report.

\section{ISP revenues and CAPEX}

Regarding the variables \textit{ISP revenues} and \textit{Capital Expenditures} ($CAPEX$), our dataset refers to the following ten major ISPs, which have been identified as the most popular in the seven countries of interest: i) NTT DoCoMo (Nippon Telegraph \& Telephone), ii) Softbank, iii) AT \& T, iv) Verizon, v) BT Group, vi) Vodafone, vii) Telecom Italia, viii) Orange (formerly France T\`el\`ecom), ix) Telef\'onica, x) Deutsche Telecom.

Tables \ref{NTT DoCoMo revenues and CAPEX}-\ref{Deutsche Telecom revenues and CAPEX} illustrate the time series of \textit{ISP revenues} and \textit{Capital Expenditures} (in millions of US dollars) of each of the aforementioned ISP companies in the considered countries, over the reference period 2008-2013. In particular, the first column contains the year, the second includes the country where the company has generated its revenues and has invested \textit{CAPEX}, which are presented in the third and in the fourth column, respectively, and, finally the fifth column illustrates the data sources\footnote{Please note that all references in this report have been accessed on July, $30^{th}$ 2015.} related to both time series of \textit{ISP revenues} and \textit{Capital Expenditures}. It is worth mentioning that the values of revenues and CAPEX generated by NTT DoCoMo, Softbank, BT Group, Vodafone, Telecom Italia, Orange, Telef\'onica and Deutsche Telecom have been converted into US dollars by employing the exchange rates acquired from \cite{Exchange2015WorldBank}.

\vspace{-1em}

\begin{center}
\renewcommand{\arraystretch}{1.1}
\begin{longtable}[h!]{K{0.15\textwidth}|K{0.15\textwidth}|K{0.15\textwidth}|K{0.15\textwidth}|K{0.15\textwidth}}
\caption{NTT DoCoMo revenues and CAPEX (in millions of US dollars)}\\
\label{NTT DoCoMo revenues and CAPEX}
\textbf{Year} & \textbf{Country} & \textbf{Annual revenues} & \textbf{CAPEX}& \textbf{Reference} \\
\hline \\[-2.8ex]\hline
2008       & Japan       &  43022.208  &  7134.348
 &  \cite{NTTDoCoMo2015NTTDoCoMo} \\ 
2009    & Japan    &  45787.244  &  7336.682 &  \cite{NTTDoCoMo2015NTTDoCoMo} \\ 
2010     & Japan       &  48133.829  &  7617.005 &  \cite{NTTDoCoMo2015NTTDoCoMo}
\\ 
2011  & Japan  &  53194.864 &  9118.810  &  \cite{NTTDoCoMo2015NTTDoCoMo} \\ 
2012     & Japan   &  56006.741 &  9442.704 &  \cite{NTTDoCoMo2015NTTDoCoMo} \\ 
2013       & Japan     &  45709.984 &  7204.287 &  \cite{NTTDoCoMo2015NTTDoCoMo} \\ 
\end{longtable}
\end{center} 

\vspace{-3.5em}

\begin{center}
\renewcommand{\arraystretch}{1.1}
\begin{longtable}[h!]{K{0.15\textwidth}|K{0.15\textwidth}|K{0.15\textwidth}|K{0.15\textwidth}|K{0.15\textwidth}}
\caption{Softbank revenues and CAPEX (in millions of US dollars)}\\
\label{Softbank revenues and CAPEX}
\textbf{Year}  & \textbf{Country} & \textbf{Annual revenues} & \textbf{CAPEX}& \textbf{Reference} \\
\hline \\[-2.8ex]\hline
2008       &  Japan       & 25854.403  &  2506.036
 &  \cite{Softbank2014Softbank} \\ 
2009 &  Japan       &  29532.403 &  2382.283 &  \cite{Softbank2015Softbank} \\ 
2010   &  Japan         &  34236.620  &  4792.459 &  \cite{Softbank2015Softbank}
\\ 
2011  &  Japan  &  40177.600 &  6478.415  &  \cite{Softbank2015Softbank} \\ 
2012   &  Japan     &  40124.991 &  9436.452 &  \cite{Softbank2015Softbank} \\ 
2013    &  Japan        &  68307.250 &  12758.970 &  \cite{Softbank2015Softbank} \\ 
\end{longtable}
\end{center}

\begin{center}
\renewcommand{\arraystretch}{1.1}
\begin{longtable}[h!]{K{0.15\textwidth}|K{0.15\textwidth}|K{0.15\textwidth}|K{0.15\textwidth}|K{0.15\textwidth}}
\caption{AT \& T revenues and CAPEX (in millions of US dollars)}\\
\label{AT and T revenues and CAPEX}
\textbf{Year} & \textbf{Country} &\textbf{Annual revenues} & \textbf{CAPEX}& \textbf{Reference} \\
\hline \\[-2.8ex]\hline
2008 & USA      &  123443  &  19631 &  \cite{ATandT2011ATandT} \\ 
2009    & USA    &  122513  &  16554 &  \cite{ATandT2014ATandT} \\ 
2010  & USA          &  124280 &  19530 &  \cite{ATandT2014ATandT}
\\ 
2011 & USA   &  126723 &  20110  &  \cite{ATandT2014ATandT} \\ 
2012    & USA    &  127434  &  19465 &  \cite{ATandT2014ATandT} \\ 
2013     & USA       &  128752 &  20944 &  \cite{ATandT2014ATandT} \\ 
\end{longtable}
\end{center}
\begin{center}
\renewcommand{\arraystretch}{1.1}
\begin{longtable}[h!]{K{0.15\textwidth}|K{0.15\textwidth}|K{0.15\textwidth}|K{0.15\textwidth}|K{0.15\textwidth}}
\caption{Verizon revenues and CAPEX (in millions of US dollars)}\\
\label{Verizon revenues and CAPEX}
\textbf{Year} & \textbf{Country} & \textbf{Annual revenues} & \textbf{CAPEX}& \textbf{Reference} \\
\hline \\[-2.8ex]\hline
2008  & USA     &  97354  &  17133 &  \cite{Verizon2011Verizon} \\ 
2009     & USA   &  107808  &  16872 &  \cite{Verizon2014Verizon} \\ 
2010    & USA        &  106565 &  16458 &  \cite{Verizon2014Verizon}
\\ 
2011   & USA &  110875 &  16244  &  \cite{Verizon2014Verizon} \\ 
2012   & USA     &  115846  &  16175 &  \cite{Verizon2014Verizon} \\ 
2013     & USA       &  120550 &  16604 &  \cite{Verizon2014Verizon} \\ 
\end{longtable}
\end{center}

\begin{center}
\renewcommand{\arraystretch}{1.1}
\begin{longtable}[h!]{K{0.15\textwidth}|K{0.15\textwidth}|K{0.15\textwidth}|K{0.15\textwidth}|K{0.15\textwidth}}
\caption{BT Group revenues and CAPEX (in millions of US dollars)}\\
\label{BT Group revenues and CAPEX}
\textbf{Year} & \textbf{Country} & \textbf{Annual revenues} & \textbf{CAPEX}& \textbf{Reference} \\
\hline \\[-2.8ex]\hline
2008  & UK    &  31476.190  &  5039.780 &  \cite{BTGroup2011BTGroup} \\ 
2009  & UK     &  26109.204  &  3759.725 &  \cite{BTGroup2011BTGroup} \\ 
2010   & UK         &  24828.439 &  3004.241 &\cite{BTGroup2011BTGroup}\\ 
2011 & UK  &  24959.936 &  3219.832  &  \cite{BTGroup2014BTGroup} \\ 
2012  & UK      &  23400.951  &  3135.727 &  \cite{BTGroup2014BTGroup} \\ 
2013  & UK          &  21978.125 &  2945.069 &  \cite{BTGroup2014BTGroup} \\ 
\end{longtable}
\end{center}

\begin{center}
\renewcommand{\arraystretch}{1.1}
\begin{longtable}[h!]{K{0.15\textwidth}|K{0.15\textwidth}|K{0.15\textwidth}|K{0.15\textwidth}|K{0.15\textwidth}}
\caption{Vodafone revenues and CAPEX (in millions of US dollars)}\\
\label{Vodafone revenues and CAPEX}
\textbf{Year} & \textbf{Country} & \textbf{Annual revenues} & \textbf{CAPEX}& \textbf{Reference} \\
\hline \\[-2.8ex]\hline
2008       &  UK & 9934.066  &  1421.032
 &  \cite{Vodafone2011Vodafone} \\ 
2009        &  UK &  8411.856  &  1211.831
 &  \cite{Vodafone2011Vodafone} \\ 
2010            &  UK &  7766.615 &  1081.374
 &\cite{Vodafone2011Vodafone}
\\ 
2011    &  UK &  8447.115 &  1144.900
  &  \cite{Vodafone2014Vodafone} \\ 
2012        &  UK &  8553.090  &  1172.855
 &  \cite{Vodafone2014Vodafone} \\ 
2013            &  UK &  8046.875 &  1134.474 &  \cite{Vodafone2014Vodafone} \\ 
2008       &  Italy & 8122.711  &  1161.924
 &  \cite{Vodafone2011Vodafone} \\ 
2009        &  Italy &  8653.666  &  1246.666
 &  \cite{Vodafone2011Vodafone} \\ 
2010            &  Italy &  9315.301 &  1297.004
 &\cite{Vodafone2011Vodafone}
\\ 
2011    &  Italy &  9169.872 &  1242.861
  &  \cite{Vodafone2014Vodafone} \\ 
2012        &  Italy &  8966.719  &  1229.575
 &  \cite{Vodafone2014Vodafone} \\ 
2013            &  Italy &  7429.688 &  1047.461 &  \cite{Vodafone2014Vodafone} \\ 
2008       &  Spain & 9272.894  &  1326.454
 &  \cite{Vodafone2011Vodafone} \\ 
2009        &  Spain &  9067.083  &  1306.224
 &  \cite{Vodafone2011Vodafone} \\ 
2010            &  Spain &  8829.985 &  1229.431
 &\cite{Vodafone2011Vodafone}
\\ 
2011    &  Spain &  8225.962 &  1114.926
  &  \cite{Vodafone2014Vodafone} \\ 
2012        &  Spain &  7548.336  &  1035.077
 &  \cite{Vodafone2014Vodafone} \\ 
2013            &  Spain &  6100.000 &  859.998 &  \cite{Vodafone2014Vodafone} \\ 
2008       &  Germany & 12575.092  &  1798.822
 &  \cite{Vodafone2011Vodafone} \\ 
2009        &  Germany &  12241.810  &  1763.582
 &  \cite{Vodafone2011Vodafone} \\ 
2010            &  Germany &  12377.125 &  1723.313
 &\cite{Vodafone2011Vodafone}
\\ 
2011    &  Germany &  12660.256 &  1715.939
  &  \cite{Vodafone2014Vodafone} \\ 
2012        &  Germany &  13047.544  &  1789.164
 &  \cite{Vodafone2014Vodafone} \\ 
2013            &  Germany &  12276.563 &  1730.790 &  \cite{Vodafone2014Vodafone} \\ 

\end{longtable}
\end{center}

\begin{table*}[h!]
\begin{center}
\renewcommand{\arraystretch}{1.1}
\caption{Orange revenues and CAPEX (in millions of US dollars)}
\label{Orange revenues and CAPEX}
\begin{tabular}{K{0.15\textwidth}|K{0.15\textwidth}|K{0.15\textwidth}|K{0.15\textwidth}|K{0.15\textwidth}}
\textbf{Year} & \textbf{Country} & \textbf{Annual revenues} & \textbf{CAPEX}& \textbf{Reference} \\
\hline
\hline
2008       &  France & 41812.865  &  6176.331
 &  \cite{Orange2009Orange} \\ 
2009        &  France &  32831.944  &  3775.674
 &  \cite{Orange2010Orange} \\ 
2010            &  France &  30871.523 &  3735.454
 &\cite{Orange2011Orange}
\\ 
2011    &  France &  31293.463 &  3974.270
  &  \cite{Orange2012Orange} \\ 
2012        &  France &  27506.427  &  3685.861
 &  \cite{Orange2013Orange} \\ 
2013            &  France &  26586.985 &  3758.300 &  \cite{Orange2014Orange} \\ 
2008       &  Spain & 4970.760  &  734.249
 &  \cite{Orange2009Orange} \\ 
2009        &  Spain &  5398.611  &  620.840
 &  \cite{Orange2010Orange} \\ 
2010            &  Spain &  5060.927 &  612.372
 &\cite{Orange2011Orange}
\\ 
2011    &  Spain &  5549.374 &  704.771
  &  \cite{Orange2012Orange} \\ 
2012        &  Spain &  5141.388  &  688.946
 &  \cite{Orange2013Orange} \\ 
2013            &  Spain &  5378.486 &  746.348 &  \cite{Orange2014Orange} \\
\end{tabular}
\end{center}
\end{table*}

\begin{table*}[h!]
\begin{center}
\renewcommand{\arraystretch}{1.1}
\caption{Telecom Italia revenues and CAPEX (in millions of US dollars)}
\label{Telecom Italia revenues and CAPEX}
\begin{tabular}{K{0.15\textwidth}|K{0.15\textwidth}|K{0.15\textwidth}|K{0.15\textwidth}|K{0.15\textwidth}}
\textbf{Year}  & \textbf{Country} & \textbf{Annual revenues} & \textbf{CAPEX}& \textbf{Reference} \\
\hline
\hline
2008 & Italy      &  33957.602  &  5347.953
 &  \cite{Telecom2010Telecom} \\ 
2009  & Italy      &  30087.500  &  4881.944 &  \cite{Telecom2011Telecom} \\ 
2010 & Italy           &  26580.132  &  4113.907 &  \cite{Telecom2012Telecom}
\\ 
2011  & Italy  &  26413.074 &  5820.584  &  \cite{Telecom2013Telecom} \\ 
2012 & Italy       &  22987.147 &  3948.586 &  \cite{Telecom2014Telecom} \\ 
2013   & Italy         &  21480.744 &  4019.920 &  \cite{Telecom2014Telecom} \\ 
\end{tabular}
\end{center}
\end{table*}

\clearpage

\begin{center}
\renewcommand{\arraystretch}{1.1}
\begin{longtable}[h!]{K{0.15\textwidth}|K{0.15\textwidth}|K{0.15\textwidth}|K{0.15\textwidth}|K{0.15\textwidth}}
\caption{Telef\'onica revenues and CAPEX (in millions of US dollars)}\\
\label{Telefonica revenues and CAPEX}
\textbf{Year} & \textbf{Country} & \textbf{Annual revenues} & \textbf{CAPEX}& \textbf{Reference} \\
\hline \\[-2.8ex]\hline
2008       &  UK & 10309.942  &  1048.246
 &  \cite{Telefonica2010Telefonica} \\ 
2009        &  UK &  9044.444  &  836.111
 &  \cite{Telefonica2010Telefonica} \\ 
2010            &  UK &  9537.748 &  949.669
 &\cite{Telefonica2011Telefonica}
\\ 
2011    &  UK &  9632.823 &  1018.081
  &  \cite{Telefonica2014Telefonica} \\ 
2012        &  UK &  9051.414  &  961.440
 &  \cite{Telefonica2014Telefonica} \\ 
2013            &  UK &  8887.118 &  1839.309 &  \cite{Telefonica2014Telefonica} \\ 
2008       &  Spain & 30464.912  &  3228.070
 &  \cite{Telefonica2010Telefonica} \\ 
2009        &  Spain &  27365.278  &  2587.500
 &  \cite{Telefonica2010Telefonica} \\ 
2010            &  Spain &  24782.781 &  2676.821
 &\cite{Telefonica2011Telefonica}
\\ 
2011    &  Spain &  24029.207 &  4050.070
  &  \cite{Telefonica2014Telefonica} \\ 
2012        &  Spain &  19275.064  &  2174.807
 &  \cite{Telefonica2014Telefonica} \\ 
2013            &  Spain &  17209.827 &  2030.544 &  \cite{Telefonica2014Telefonica} \\ 
2008       &  Germany & 5255.848  &  1350.877  &  \cite{Telefonica2010Telefonica} \\ 
2009        &  Germany &  5202.778  &  1105.556 &  \cite{Telefonica2010Telefonica} \\ 
2010            &  Germany &  6392.053 & 2724.503 &\cite{Telefonica2011Telefonica}
\\ 
2011    &  Germany &  7002.782 &  776.078 &  \cite{Telefonica2014Telefonica} \\ 
2012        &  Germany &  6700.514  &  782.776
 &  \cite{Telefonica2014Telefonica} \\ 
2013            &  Germany &  6525.896 &  884.462 &  \cite{Telefonica2014Telefonica} \\ 
\end{longtable}
\end{center}

\begin{center}
\renewcommand{\arraystretch}{1.1}
\begin{longtable}[h!]{K{0.15\textwidth}|K{0.15\textwidth}|K{0.15\textwidth}|K{0.15\textwidth}|K{0.15\textwidth}}
\caption{Deutsche Telecom revenues and CAPEX (in millions of US dollars)}\\
\label{Deutsche Telecom revenues and CAPEX}
\textbf{Year} & \textbf{Country} & \textbf{Annual revenues} & \textbf{CAPEX}& \textbf{Reference} \\
\hline \hline \\[-2.8ex]\hline
2008       &  USA & 21866.959  &  3713.450
 &  \cite{DeutscheTelecom2011DeutscheTelecom} \\ 
2009        &  USA &  21487.500  &  3702.778
 &  \cite{DeutscheTelecom2011DeutscheTelecom} \\ 
2010            &  USA &  21307.285 &  2809.272
 &\cite{DeutscheTelecom2011DeutscheTelecom}
\\ 
2011    &  USA &  20599.444 &  2730.181
  &  \cite{DeutscheTelecom2014DeutscheTelecom} \\ 
2012        &  USA &  19757.069  &  3290.488
 &  \cite{DeutscheTelecom2014DeutscheTelecom} \\ 
2013            &  USA &  24642.762 &  4354.582 &  \cite{DeutscheTelecom2014DeutscheTelecom} \\
2008       &  Germany & 38596.491  &  4441.520
 &  \cite{DeutscheTelecom2011DeutscheTelecom} \\ 
2009        &  Germany &  35309.722  &  4386.111
 &  \cite{DeutscheTelecom2011DeutscheTelecom} \\ 
2010            &  Germany &  33304.636 &  6311.258
 &\cite{DeutscheTelecom2011DeutscheTelecom}
\\ 
2011    &  Germany &  32275.382 &  4876.217
  &  \cite{DeutscheTelecom2014DeutscheTelecom} \\ 
2012        &  Germany &  29223.650  &  4393.316
 &  \cite{DeutscheTelecom2014DeutscheTelecom} \\ 
2013            &  Germany &  29794.157 &  4529.880 &  \cite{DeutscheTelecom2014DeutscheTelecom} \\ 
\end{longtable}
\end{center}

Finally, Table \ref{ISP revenues and CAPEX} illustrates the aggregate series of \textit{ISP revenues} and \textit{CAPEX} (in millions of US dollars) in the considered countries, over the reference period 2008-2013, obtained by summing the \textit{ISP revenues} and \textit{Capital Expenditures} available for every single ISP company.

\begin{center}
\renewcommand{\arraystretch}{1.1}
\begin{longtable}[h!]{K{0.20\textwidth}|K{0.20\textwidth}|K{0.20\textwidth}|K{0.20\textwidth}}
\caption{ISP revenues and CAPEX (in millions of US dollars)}\\
\label{ISP revenues and CAPEX}
\textbf{Year} & \textbf{Country} & \textbf{Annual revenues} & \textbf{CAPEX} \\
\hline \\[-2.8ex]\hline
2008 & Japan  & 68876.61 & 9640.38\\ 
2009 & Japan       & 75319.65 & 9718.97\\ 
2010 & Japan       & 82370.45 & 12409.46\\ 
2011 & Japan       & 93372.46 & 15597.22\\ 
2012 & Japan       & 96131.73 & 18879.16 \\ 
2013 & Japan       & 114017.23 & 19963.26\\ 
2008 & USA   & 242663.96 & 40477.45 \\ 
2009 & USA  & 251808.50 & 37128.78 \\ 
2010 &  USA    & 252152.28 & 38797.27 \\ 
2011 &  USA    & 258197.44 & 39084.18 \\ 
2012 & USA            & 263037.07 & 38930.49 \\ 
2013 & USA            & 273944.76 & 41902.58 \\ 
2008 & UK          & 51720.20 & 7509.06 \\ 
2009 & UK          & 43565.51 & 5807.67 \\ 
2010 & UK          & 42132.80 & 5035.28 \\ 
2011 & UK          & 43039.87 & 5382.81 \\ 
2012 & UK          & 41005.46 & 5270.02 \\ 
2013 & UK          & 38912.12 & 5918.85 \\
2008 & France        & 41812.87 & 6176.33 \\ 
2009 & France        & 32831.94 & 3775.67 \\ 
2010 & France        & 30871.52 & 3735.45 \\ 
2011 & France        & 31293.46 & 3974.27 \\ 
2012 & France        & 27506.43 & 3685.86 \\ 
2013 & France        & 26586.99 & 3758.30 \\
2008 & Italy        & 42080.31 & 6509.88 \\ 
2009 & Italy        & 38741.17 & 6128.61 \\ 
2010 & Italy        & 35895.43 & 5410.91 \\ 
2011 & Italy        & 35582.95 & 7063.45 \\ 
2012 &  Italy        & 31953.87 &  5178.16 \\ 
2013 & Italy        & 28910.43 & 5067.38 \\
2008 & Spain        & 44708.57 &  5288.77 \\ 
2009 & Spain        & 41830.97 & 4514.56 \\ 
2010 & Spain        & 38673.69 & 4518.62 \\ 
2011 & Spain        & 37804.54 &  5869.77 \\ 
2012 & Spain        & 31964.79 & 3898.83 \\
2013 & Spain        & 28688.31 & 3636.89 \\ 
2008 & Germany        & 56427.43 & 7591.22 \\ 
2009 & Germany        & 52754.31 & 7255.25 \\ 
2010 & Germany        & 52073.81 & 10759.07 \\ 
2011 & Germany        & 51938.42 & 7368.23 \\ 
2012 &  Germany        & 48971.71 & 6965.26 \\
2013 &  Germany        & 48596.62 & 7145.13 \\
\end{longtable}
\end{center}

\section{OTT revenues}

Among the numerous OTT suppliers at play, our dataset focuses on the most popular ones, who offer similar services (e.g., voice and instant messaging) to those of the ISPs:
i) Skype, ii) Facebook, iii) Whatsapp. Therefore, the \textit{OTT revenues} refer to the revenues generated by the three aforementioned OTT providers in the considered countries over the reference period 2008-2013.

\subsection{Skype}

As the Skype revenues per country are not available, a rough estimation which employs the \textit{Skype total users}, the \textit{Skype users by country} and the \textit{Skype total revenues} has been made. 

Table \ref{Skype total users} illustrates the \textit{Skype total users} over the reference period 2008-2013. In particular, the first column contains the year, the second column includes the number of \textit{Skype total users} in millions of people, while the third one illustrates the sources from which data has been obtained.

\begin{table*}[h!]
\begin{center}
\renewcommand{\arraystretch}{1.1}
\caption{Skype total users (in millions of people)}
\label{Skype total users}
\begin{tabular}{K{0.30\textwidth}|K{0.30\textwidth}|K{0.30\textwidth}}
\textbf{Year} & \textbf{Users} & \textbf{Reference} \\
\hline
\hline
2008       &  75 &  \cite{Fujimoto2013Marketing} \\ 
2009        &  105 &  \cite{Fujimoto2013Marketing} \\ 
2010            &  145 &  \cite{Fujimoto2013Marketing}
\\ 
2011    &  200 &  \cite{Mercier2013Skype} \\ 
2012        &  280 &  \cite{Paraskelidis2013Evaluating} \\ 
2013            &  300 &  \cite{Warren2014Viber} \\ 
\end{tabular}
\end{center}
\end{table*}

Table \ref{Skype users by country} illustrates the \textit{Skype users by country}. Specifically, the first and the second columns contain the reference year and country, respectively, while the third and the fourth columns illustrate the country smartphone penetration rate, a proxy of the Skype penetration rate by country, based on the assumption that smartphone users are also Skype users, and the reference from which the data have been acquired. It is worth mentioning that due to the absence of available data for 2008 and 2009, the corresponding values (in red) in Table \ref{Skype users by country} have been obtained by interpolation\footnote{A comparison with the available data gives errors lower than 20\% (in particular, lower than 10\% in the 95.45\% of the cases and lower than 5\% in the 77.27\% of the cases).}, while the blue values in the table correspond to forecasts made by \cite{eMarketer2013Three}.

\begin{center}
\renewcommand{\arraystretch}{1.1}
\begin{longtable}[h!]{K{0.20\textwidth}|K{0.20\textwidth}|K{0.20\textwidth}|K{0.20\textwidth}}
\caption{Skype users by country}\\
\label{Skype users by country}
\textbf{Year} & \textbf{Country} & \textbf{Users} & \textbf{Reference} \\
\hline \\[-2.8ex]\hline
2008 & Japan  & \textcolor{red}{0.007}  &  - \\ 
2009 & Japan       & \textcolor{red}{0.021}  &  - \\ 
2010 & Japan       & 0.065  &  \cite{eMarketer2013Three} \\
2011 & Japan       & 0.180 &  \cite{eMarketer2013Three} \\
2012 & Japan       & 0.330 &  \cite{eMarketer2013Three} \\  
2013 & Japan       & \textcolor{blue}{0.490} &  \cite{eMarketer2013Three} \\ 
2008 & USA   & \textcolor{red}{0.093}  &  - \\ 
2009 & USA  & \textcolor{red}{0.188}  &  - \\ 
2010 &  USA    & 0.269  &  \cite{eMarketer2013Three} \\ 
2011 &  USA    & 0.392 &  \cite{eMarketer2013Three} \\ 
2012 & USA        & 0.477 &  \cite{eMarketer2013Three} \\  
2013 & USA        & \textcolor{blue}{0.555} &  \cite{eMarketer2013Three} \\ 
2008 & UK          & \textcolor{red}{0.003}  &  - \\  
2009 & UK          & \textcolor{red}{0.106}  &  - \\  
2010 & UK          & 0.200  &  \cite{eMarketer2013Three} \\ 
2011 & UK          & 0.300 &  \cite{eMarketer2013Three} \\ 
2012 & UK          & 0.368 &  \cite{eMarketer2013Three} \\
2013 & UK          &  \textcolor{blue}{0.455} &  \cite{eMarketer2013Three} \\
2008 & France        & \textcolor{red}{0.099}  &  - \\ 
2009 & France        & \textcolor{red}{0.124}  &  - \\ 
2010 & France      & 0.170  &  \cite{eMarketer2013Three} \\ 
2011 & France      & 0.245 &  \cite{eMarketer2013Three} \\ 
2012 & France      & 0.330 &  \cite{eMarketer2013Three} \\
2013 & France      &  \textcolor{blue}{0.450} &  \cite{eMarketer2013Three} \\ 
2008 & Italy        & \textcolor{red}{0.007}  &  - \\  
2009 & Italy        & \textcolor{red}{0.072}  &  - \\  
2010 & Italy      &  0.130  &  \cite{eMarketer2013Three} \\
2011 & Italy      & 0.240 &  \cite{eMarketer2013Three} \\ 
2012 &  Italy     & 0.314 &  \cite{eMarketer2013Three} \\ 
2013 & Italy      &  \textcolor{blue}{0.410} &  \cite{eMarketer2013Three} \\ 
2008 & Spain        & \textcolor{red}{0.039}  &  - \\
2009 & Spain        & \textcolor{red}{0.077}  &  - \\ 
2010 & Spain     & 0.130  &  \cite{eMarketer2013Three} \\ 
2011 & Spain     &  0.200 &  \cite{eMarketer2013Three} \\
2012 & Spain        & 0.280 &  \cite{eMarketer2013Three} \\ 
2013 & Spain     &   \textcolor{blue}{0.380} &  \cite{eMarketer2013Three} \\ 
2008 & Germany        & \textcolor{red}{0.003}  &  - \\ 
2009 & Germany        & \textcolor{red}{0.049}  &  - \\ 
2010 & Germany    & 0.100  &  \cite{eMarketer2013Three} \\ 
2011 & Germany    & 0.185 &  \cite{eMarketer2013Three} \\
2012 &  Germany   & 0.270 &  \cite{eMarketer2013Three} \\ 
2013 &  Germany   &  \textcolor{blue}{0.360} &  \cite{eMarketer2013Three} \\ 
\end{longtable}
\end{center}

\clearpage

Table \ref{Skype total revenues} illustrates the \textit{Skype total revenues}. Specifically, the first column contains the reference year, while the second and the third columns illustrate the total annual revenues, and the reference from which the data have been acquired. Due to the absence of available data about the Skype annual revenues in 2012, the respective value (in red) has been obtained by interpolation, while, regarding the year 2011, the value in blue corresponds to an estimation made by \cite{furrer2016corporate}.

\begin{table*}[h!]
\begin{center}
\renewcommand{\arraystretch}{1.1}
\caption{Skype total revenues (in millions of US dollars)}
\label{Skype total revenues}
\begin{tabular}{K{0.30\textwidth}|K{0.30\textwidth}|K{0.30\textwidth}}
\textbf{Year} & \textbf{Annual revenues} & \textbf{Reference} \\
\hline
\hline
2008       &  551.36  &  \cite{Statista2016Skype} \\ 
2009        &  718.90  &  \cite{Statista2016Skype} \\ 
2010            &  859.82  &  \cite{Statista2016Skype}
\\ 
2011    &  \textcolor{blue}{1000.00} &  \cite{furrer2016corporate} \\ 
2012        &  \textcolor{red}{1478.33} &  - \\ 
2013            &  2000.00 &  \cite{Bass2013Microsoft} \\ 
\end{tabular}
\end{center}
\end{table*}

Finally, Table \ref{Skype revenues by country} illustrates the \textit{Skype revenues by country}. In particular, the first and the second columns contain the reference year and country, respectively, while the third one illustrates the total annual revenues by country approximated by the following formula:

\begin{gather}
\begin{aligned}\label{eq}
Skype \ revenues \ by \ country \approx \frac{Skype \ total \ revenues * Skype \ users \ by \ country}{Skype \ total \ users}.
\end{aligned}
\end{gather}

\begin{center}
\renewcommand{\arraystretch}{1.1}
\begin{longtable}[h!]{K{0.30\textwidth}|K{0.30\textwidth}|K{0.30\textwidth}}
\caption{Skype revenues by country (in millions of US dollars)}\\
\label{Skype revenues by country}
\textbf{Year} & \textbf{Country} & \textbf{Annual revenues} \\
\hline \\[-2.8ex]\hline
2008 & Japan &  0.051  \\ 
2009 & Japan      &  0.146 \\ 
2010 & Japan       &  0.385 \\
2011 & Japan        &  0.900 \\ 
2012 & Japan       &  1.742 \\
2013 & Japan       &  3.267 \\ 
2008 & USA   &  0.685  \\ 
2009 & USA &  1.284 \\
2010 &  USA      &  1.595 \\ 
2011 &  USA    &  1.960 \\ 
2012 & USA           &  2.518 \\ 
2013 & USA            &  3.700 \\ 
2008 & UK           &  0.025  \\ 
2009 & UK          &  0.727 \\  
2010 & UK          &  1.186 \\ 
2011 & UK         &  1.500 \\ 
2012 & UK           &  1.943 \\  
2013 & UK          &  3.033 \\ 
2008 & France         &  0.726  \\ 
2009 & France        &  0.847 \\ 
2010 & France       &  1.008 \\ 
2011 & France         &  1.225 \\  
2012 & France       &  1.742 \\ 
2013 & France       &  3.000 \\ 
2008 & Italy        &  0.053  \\ 
2009 & Italy        &  0.490 \\ 
2010 & Italy         &  0.771 \\ 
2011 & Italy         &  1.200 \\
2012 &  Italy         &  1.658 \\
2013 & Italy          &  2.733 \\ 
2008 & Spain         &  0.290  \\ 
2009 & Spain         & 0.531 \\
2010 & Spain        &  0.771 \\ 
2011 & Spain        &  1.000 \\ 
2012 & Spain        &  1.478 \\ 
2013 & Spain         &  2.533 \\ 
2008 & Germany         &  0.023  \\ 
2009 & Germany          & 0.336 \\  
2010 & Germany      &  0.593 \\
2011 & Germany         &  0.925 \\ 
2012 &  Germany          &  1.426 \\ 
2013 &  Germany        &  2.400 \\ 
\end{longtable}
\end{center}

\subsection{Facebook}

As the Facebook revenue per country is not available for each of the considered countries, in the absence of data, a rough estimation which employs the \textit{Facebook total users}, the \textit{Facebook users by country} and the \textit{Facebook total revenues} has been made. 

Table \ref{Facebook total users} illustrates the \textit{Facebook total users} over the reference period 2008-2013. In particular, the first column contains the year, the second includes the number of \textit{Facebook total users} (in millions of people), while the third one illustrates the sources from which data have been obtained.

\begin{table*}[h!]
\begin{center}
\renewcommand{\arraystretch}{1.1}
\caption{Facebook total users (in millions of people)}
\label{Facebook total users}
\begin{tabular}{K{0.30\textwidth}|K{0.30\textwidth}|K{0.30\textwidth}}
\textbf{Year} & \textbf{Users} & \textbf{Reference} \\
\hline
\hline
2008       &  145 &  \cite{Sedghi2014Facebook} \\ 
2009        &  360 &  \cite{Sedghi2014Facebook} \\ 
2010            &  608 &  \cite{Sedghi2014Facebook}
\\ 
2011    &  845 &  \cite{Sedghi2014Facebook} \\ 
2012        &  1056 &  \cite{Sedghi2014Facebook} \\ 
2013            &  1230 &  \cite{Sedghi2014Facebook} \\ 
\end{tabular}
\end{center}
\end{table*}

Table \ref{Facebook users by country} illustrates the \textit{Facebook users by country}. Specifically, the first and the second columns contain the reference year and country, respectively, while the third and the fourth columns illustrate the number of \textit{Facebook users by country} and the reference from which the data have been acquired. As it is possible to notice by Table \ref{Facebook users by country}, there is no value related to Facebook users in Japan in 2008. This is due to the fact that the japanese version of Facebook has been launched in 2008 \cite{Tabuchi2011Facebook,hamada2012facebook}, therefore there are no available data before 2009. Moreover, it is worth mentioning that due to the absence of available data about UK Facebook users in 2010, the respective value (in red) in Table \ref{Facebook users by country} has been obtained by interpolation, while the blue values correspond to adjustments made by interpolation in the case in which the collected values were related to the months of July and September and not to the end of the year such as all the other values.

\begin{center}
\renewcommand{\arraystretch}{1.1}
\begin{longtable}[h!]{K{0.20\textwidth}|K{0.20\textwidth}|K{0.20\textwidth}|K{0.20\textwidth}}
\caption{Facebook users by country (in millions of people)}\\
\label{Facebook users by country}
\textbf{Year} & \textbf{Country} & \textbf{Users} & \textbf{Reference} \\
\hline \\[-2.8ex]\hline
2008 & Japan   &  -  &  - \\ 
2009 & Japan       &  1.00  &  \cite{Darwell2012Facebook} \\ 
2010 & Japan      &  6.00  &  \cite{Bloomberg2012Japan} \\ 
2011 & Japan       &  13.50 &  \cite{Bloomberg2012Japan} \\ 
2012 & Japan       &  23.20 &  \cite{Statista2014Number} \\ 
2013 & Japan       &  25.30 &  \cite{Statista2014Number} \\
2008 & USA   &  33.00  &  \cite{SocialTimes2008Latest} \\ 
2009 & USA    &  103.00  &  \cite{SocialTimes2010December} \\ 
2010 &  USA   &  138.60 &  \cite{Eldon2010Facebook} \\ 
2011 &  USA     &  149.40 &  \cite{Lee2011Facebook} \\ 
2012 & USA        &  169.00 &  \cite{Kiss2013Facebook} \\ 
2013 & USA        &  180.00 &  \cite{Saul2014MillionTeens} \\ 
2008 & UK         &  \textcolor{blue}{12.00}  &  \cite{WikidotSocial} \\  
2009 & UK        &  \textcolor{blue}{18.46}  &  \cite{Burcher2010Facebook} \\
2010 & UK         &  \textcolor{red}{23.41}  &  - \\ 
2011 & UK        &  25.60 &  \cite{eMarketer2013Emerging} \\  
2012 & UK       &  28.30 &  \cite{eMarketer2013Emerging} \\ 
2013 & UK        &  29.90 &  \cite{eMarketer2013Emerging} \\ 
2008 & France  &  6.54  &  \cite{SocialTimes2009Bonjour} \\ 
2009 & France    &  \textcolor{blue}{14.45}  &  \cite{Burcher2010Facebook} \\ 
2010 & France     &  22.00 &  \cite{Rao2011Facebook} \\ 
2011 & France     &  23.00 &  \cite{Yeates2012UK} \\ 
2012 & France     &  25.62 &  \cite{InternetInternet} \\ 
2013 & France     &  26.00 &  \cite{Constine2013Facebook} \\  
2008 & Italy    &  4.90  & \cite{Mysore2008FacebookItaly} \\ 
2009 & Italy   &  \textcolor{blue}{12.71}  &  \cite{Burcher2010Facebook} \\ 
2010 & Italy     &  \textcolor{blue}{18.19}  &  \cite{Burcher2010Facebook} \\ 
2011 & Italy      &  \textcolor{blue}{21.70} &  \cite{YungHui2012Billion} \\  
2012 &  Italy      &  23.20 &  \cite{InternetInternet} \\
2013 & Italy      &  23.00 &  \cite{Constine2013Facebook} \\  
2008 & Spain      &  2.30  &  \cite{Mysore2008Footprint} \\ 
2009 & Spain    &  11.50  &  \cite{Parkinson2011Tuenti} \\
2010 & Spain     &  15.00 &  \cite{Parkinson2011Tuenti} \\ 
2011 & Spain      &  16.00 &  \cite{LATEVAWEBHowMany} \\ 
2012 & Spain        &  17.59 &  \cite{InternetInternet} \\  
2013 & Spain       &  18.00 &  \cite{Constine2013Facebook} \\ 
2008 & Germany       &  1.20  &  \cite{Mysore2008Germany} \\ 
2009 & Germany   &  \textcolor{blue}{9.48}  &  \cite{Burcher2010Facebook} \\ 
2010 & Germany  &  18.00  & \cite{Socialnomics2013Facebook} \\  
2011 & Germany     &  22.00 &  \cite{Yeates2012UK} \\
2012 &  Germany   &  25.33 &  \cite{InternetInternet} \\
2013 &  Germany    &  25.00 &  \cite{Constine2013Facebook} \\
\end{longtable}
\end{center}

\vspace{3em}

Table \ref{Facebook total revenues} illustrates the \textit{Facebook total revenues}. Specifically, the first column contains the reference year, while the second and the third columns illustrate the total annual revenues and the reference from which the data have been acquired, namely the company's reports.

\begin{table*}[h!]
\begin{center}
\renewcommand{\arraystretch}{1.1}
\caption{Facebook total revenues (in millions of US dollars)}
\label{Facebook total revenues}
\begin{tabular}{K{0.30\textwidth}|K{0.30\textwidth}|K{0.30\textwidth}}
\textbf{Year} & \textbf{Annual revenues} & \textbf{Reference} \\
\hline
\hline
2008       &  272  &  \cite{Facebook2013Facebook} \\ 
2009        &  777  &  \cite{Facebook2014Facebook} \\ 
2010            &  1974  &  \cite{Facebook2014Facebook}
\\ 
2011    &  3711 &  \cite{Facebook2014Facebook} \\ 
2012        &  5089 &  \cite{Facebook2014Facebook} \\ 
2013            &  7872 &  \cite{Facebook2014Facebook} \\ 
\end{tabular}
\end{center}
\end{table*}

Finally, Table \ref{Facebook revenues by country} illustrates the \textit{Facebook revenues by country}. In particular, the first and the second columns contain the reference year and country, respectively, while the third one illustrates the total annual revenues by country approximated by the following formula:

\begin{gather}
\scalebox{0.9}{$\label{eq}
\begin{aligned}
Facebook \ revenues \ by \ country \approx \frac{Facebook \ total \ revenues * Facebook \ users \ by \ country}{Facebook \ total \ users}.
\end{aligned}$}
\end{gather}

In the specific case of the USA, the available Facebook revenues generated in the USA have been collected from the references reported in the fourth column of Table \ref{Facebook revenues by country}. However, due to the absence of available data for the years 2008 and 2009, the related values, represented in red in Table \ref{Facebook revenues by country}, have been obtained by interpolation.

\begin{center}
\renewcommand{\arraystretch}{1.1}
\begin{longtable}[h!]{K{0.20\textwidth}|K{0.20\textwidth}|K{0.20\textwidth}|K{0.20\textwidth}}
\caption{Facebook revenues by country (in millions of US dollars)}\\
\label{Facebook revenues by country}
\textbf{Year} & \textbf{Country} & \textbf{Annual revenues} & \textbf{Reference} \\
\hline \\[-2.8ex]\hline
2008 & Japan  &  -  \\ 
2009 & Japan      &  2.16 \\ 
2010 & Japan       &  19.48 \\ 
2011 & Japan      &  59.29 \\ 
2012 & Japan       &  111.80 \\ 
2013 & Japan        &  161.92 \\ 
2008 & USA    &  \textcolor{red}{207.15} &  - \\ 
2009 & USA    &  \textcolor{red}{688.75} &  - \\
2010 &  USA       &  1223.00 &  \cite{Facebook2013Facebook} \\
2011 &  USA    &  2067.00 &  \cite{Facebook2014Facebook} \\
2012 & USA       &  2578.00 &  \cite{Facebook2014Facebook} \\
2013 & USA      &  3613.00 &  \cite{Facebook2014Facebook} \\
2008 & UK            &  22.51  \\ 
2009 & UK         &  39.85 \\  
2010 & UK      &  76.01 \\ 
2011 & UK         &  112.43 \\ 
2012 & UK             &  136.38 \\ 
2013 & UK               &  191.36 \\ 
2008 & France          &  12.27  \\ 
2009 & France       &  31.19 \\ 
2010 & France          &  71.43 \\ 
2011 & France       &  101.01 \\ 
2012 & France        &  123.47 \\ 
2013 & France       &  166.40 \\ 
2008 & Italy       &  9.19  \\ 
2009 & Italy      & 27.43 \\  
2010 & Italy        &  59.06 \\ 
2011 & Italy         &  95.30 \\ 
2012 &  Italy         &  111.80 \\ 
2013 & Italy           &  147.20 \\ 
2008 & Spain          &  4.31  \\  
2009 & Spain    &  24.82 \\ 
2010 & Spain        &  48.70 \\ 
2011 & Spain       &  70.27 \\ 
2012 & Spain      &  84.77 \\ 
2013 & Spain       &  115.20 \\
2008 & Germany        &  2.25  \\ 
2009 & Germany          &  20.46 \\ 
2010 & Germany           &  58.44 \\ 
2011 & Germany        &  96.62 \\ 
2012 &  Germany        &  122.07 \\ 
2013 &  Germany        &  160.00 \\ 
\end{longtable}
\end{center}

\subsection{WhatsApp}

Similarly to Skype, as the WhatsApp revenue per country is not available, a rough estimation which employs the \textit{WhatsApp total users}, the \textit{WhatsApp users by country} and the \textit{WhatsApp total revenues} has been made. 

Table \ref{WhatsApp total users} illustrates the \textit{WhatsApp total users} over the reference period 2008-2013. In particular, the first column contains the year, the second includes the number of \textit{WhatsApp total users} in millions of people, while the third one illustrates the sources from which data has been obtained. As it can be noticed in Table \ref{WhatsApp total users}, there is no value related to WhatsApp users in 2008. This is due to the fact that WhatsApp has been established in 2009 \cite{Flore2015Things,Satariano2014WhatsApp}. Moreover, due to the absence of available data for the year 2011, the related value, represented in red in Table \ref{WhatsApp total users}, has been obtained by interpolation.

\begin{table*}[h!]
\begin{center}
\renewcommand{\arraystretch}{1.1}
\caption{WhatsApp total users (in millions of people)}
\label{WhatsApp total users}
\begin{tabular}{K{0.28\textwidth}|K{0.28\textwidth}|K{0.28\textwidth}}
\textbf{Year} & \textbf{Users} & \textbf{Reference} \\
\hline
\hline
2008       &  - &  - \\ 
2009        &  1.00 &  \cite{Bradshaw2011WhatsApp} \\ 
2010            &  10.00 &  \cite{Bradshaw2011WhatsApp}
\\ 
2011    &  \textcolor{red}{106.50} &  - \\ 
2012        &  250.00 &  \cite{Lomas2013Skype} \\ 
2013            &  400.00 &  \cite{Rowan2014WhatsApp} \\ 
\end{tabular}
\end{center}
\end{table*}

Table \ref{WhatsApp users by country} illustrates the \textit{WhatsApp users by country}. Specifically, the first and the second columns contain the reference year and country, respectively, while the third and the fourth columns illustrate the country smartphone penetration rate, a proxy of the WhatsApp penetration rate by country, based on the assumption that smartphone users are also WhatsApp users, and the reference from which the data have been acquired. It is worth mentioning that due to the absence of available data for 2008 and 2009, the related values, represented in red in Table \ref{WhatsApp users by country}, have been obtained by interpolation, while the blue values in the table correspond to forecasts made by \cite{eMarketer2013Three}.

\vspace{-1em}

\begin{center}
\renewcommand{\arraystretch}{1.1}
\begin{longtable}[h!]{K{0.20\textwidth}|K{0.20\textwidth}|K{0.20\textwidth}|K{0.20\textwidth}}
\caption{WhatsApp users by country}\\
\label{WhatsApp users by country}
\textbf{Year} & \textbf{Country} & \textbf{Users} & \textbf{Reference} \\
\hline \\[-2.8ex]\hline
2008 & Japan   &  -  &  - \\ 
2009 & Japan       & \textcolor{red}{0.021}  &  - \\ 
2010 & Japan  &  0.065  &  \cite{eMarketer2013Three} \\ 
2011 & Japan        &  0.180 &  \cite{eMarketer2013Three} \\
2012 & Japan      &  0.330 &  \cite{eMarketer2013Three} \\  
2013 & Japan       &  \textcolor{blue}{0.490} &  \cite{eMarketer2013Three} \\
2008 & USA    &  -  &  - \\ 
2009 & USA    &  \textcolor{red}{0.188}  &  - \\ 
2010 &  USA &  0.269  &  \cite{eMarketer2013Three} \\ 
2011 &  USA      &  0.392 &  \cite{eMarketer2013Three} \\ 
2012 & USA         &  0.477 &  \cite{eMarketer2013Three} \\ 
2013 & USA        &  \textcolor{blue}{0.555} &  \cite{eMarketer2013Three} \\  
2008 & UK        &  -  &  - \\  
2009 & UK         &  \textcolor{red}{0.106}  &  - \\
2010 & UK   &  0.200  &  \cite{eMarketer2013Three} \\ 
2011 & UK        &  0.300 &  \cite{eMarketer2013Three} \\ 
2012 & UK        &  0.368 &  \cite{eMarketer2013Three} \\  
2013 & UK         &  \textcolor{blue}{0.455} &  \cite{eMarketer2013Three} \\ 
2008 & France  &  -  &  - \\  
2009 & France   &  \textcolor{red}{0.124}  &  - \\ 
2010 & France   &  0.170  &  \cite{eMarketer2013Three} \\ 
2011 & France    &  0.245 &  \cite{eMarketer2013Three} \\ 
2012 & France      &  0.330 &  \cite{eMarketer2013Three} \\ 
2013 & France     &  \textcolor{blue}{0.450} &  \cite{eMarketer2013Three} \\   
2008 & Italy    &  - &  - \\ 
2009 & Italy     &  \textcolor{red}{0.072}  &  - \\ 
2010 & Italy      &  0.130  &  \cite{eMarketer2013Three} \\ 
2011 & Italy     &  0.240 &  \cite{eMarketer2013Three} \\ 
2012 &  Italy    &  0.314 &  \cite{eMarketer2013Three} \\ 
2013 & Italy       &   \textcolor{blue}{0.410} &  \cite{eMarketer2013Three} \\  
2008 & Spain      &  -  &  - \\  
2009 & Spain    &  \textcolor{red}{0.077}  &  - \\ 
2010 & Spain     &  0.130  &  \cite{eMarketer2013Three} \\  
2011 & Spain       &  0.200 &  \cite{eMarketer2013Three} \\ 
2012 & Spain       &  0.280 &  \cite{eMarketer2013Three} \\   
2013 & Spain      &   \textcolor{blue}{0.380} &  \cite{eMarketer2013Three} \\  
2008 & Germany      &  -  &  - \\ 
2009 & Germany   &  \textcolor{red}{0.049}  &  - \\ 
2010 & Germany  &  0.100  &  \cite{eMarketer2013Three}\\ 
2011 & Germany   &  0.185 &  \cite{eMarketer2013Three} \\ 
2012 &  Germany   &  0.270 &  \cite{eMarketer2013Three} \\ 
2013 &  Germany    &  \textcolor{blue}{0.360} &  \cite{eMarketer2013Three} \\ 
\end{longtable}
\end{center}

Table \ref{WhatsApp total revenues} illustrates the \textit{WhatsApp total revenues}. Specifically, the first column contains the reference year, while the second and the third columns illustrate the total annual revenues and the reference from which the data have been acquired. Due to the absence of available data about the WhatsApp annual revenues in 2009, 2010 and 2011, the related values (in red) have been obtained by interpolation.

\begin{table*}[h!]
\begin{center}
\renewcommand{\arraystretch}{1.1}
\caption{WhatsApp total revenues (in millions of US dollars)}
\label{WhatsApp total revenues}
\begin{tabular}{K{0.28\textwidth}|K{0.28\textwidth}|K{0.28\textwidth}}
\textbf{Year} & \textbf{Annual revenues} & \textbf{Reference} \\
\hline
\hline
2008       &  -  &  - \\ 
2009        & \textcolor{red}{0.000037}  &  - \\ 
2010            &  \textcolor{red}{0.004549}  &  -
\\ 
2011    &  \textcolor{red}{0.641049} &  - \\ 
2012        &  3.820000 &  \cite{Statista2014Annual} \\ 
2013            &  10.210000 &  \cite{Statista2014Annual} \\ 
\end{tabular}
\end{center}
\end{table*}

Finally, Table \ref{WhatsApp revenues by country} illustrates the WhatsApp revenues by country. In particular, the first and the second columns contain the reference year and country, respectively, while the third one illustrate the total annual revenues by country approximated by the following formula:

\begin{gather}\label{eq}
\scalebox{0.9}{$
   \begin{aligned}
WhatsApp \ revenues \ by \ country \approx \frac{WhatsApp \ total \ revenues * WhatsApp \ users \ by \ country}{WhatsApp \ total \ users}
\end{aligned}$}
\end{gather}

\begin{center}
\renewcommand{\arraystretch}{1.1}
\begin{longtable}[h!]{K{0.28\textwidth}|K{0.28\textwidth}|K{0.28\textwidth}}
\caption{WhatsApp revenues by country (in millions of US dollars)}\\
\label{WhatsApp revenues by country}
\textbf{Year} & \textbf{Country} & \textbf{Annual revenues} \\
\hline \\[-2.8ex]\hline
2008 & Japan  &  -  \\ 
2009 & Japan        &  0.000001 \\ 
2010 & Japan        &  0.000030 \\ 
2011 & Japan       &  0.001083 \\ 
2012 & Japan        &  0.005042 \\ 
2013 & Japan       &  0.012507 \\ 
2008 & USA   &  -  \\ 
2009 & USA &  0.000007 \\ 
2010 &  USA        &  0.000122 \\ 
2011 &  USA    &   0.002360 \\ 
2012 & USA           &  0.007289 \\ 
2013 & USA          &  0.014166 \\ 
2008 & UK         &  -  \\  
2009 & UK            &  0.000004 \\ 
2010 & UK           &  0.000091 \\ 
2011 & UK          &  0.001806 \\ 
2012 & UK           &  0.005623 \\ 
2013 & UK            &  0.011614 \\ 
2008 & France         &  -  \\
2009 & France        &  0.000005 \\ 
2010 & France     &  0.000077 \\ 
2011 & France         &  0.001475 \\ 
2012 & France        &  0.005042 \\
2013 & France       &  0.011486\\ 
2008 & Italy        &  -  \\ 
2009 & Italy     &  0.000003 \\ 
2010 & Italy       &  0.000059 \\ 
2011 & Italy          &  0.001445 \\ 
2012 &  Italy             &  0.004798 \\ 
2013 & Italy            &  0.010465 \\  
2008 & Spain         &  -  \\ 
2009 & Spain        & 0,000003 \\ 
2010 & Spain           &  0,000059 \\ 
2011 & Spain       &  0,001204 \\
2012 & Spain       &  0,004278 \\ 
2013 & Spain      &  0,009700 \\ 
2008 & Germany         &  -  \\ 
2009 & Germany       & 0.000002 \\ 
2010 & Germany      &  0.000045 \\ 
2011 & Germany          &  0.001114 \\ 
2012 &  Germany       &  0.004126 \\ 
2013 &  Germany          &  0.009189 \\ 
\end{longtable}
\end{center}

\subsection{OTT revenues: the aggregate variable}

With regard to the aggregate variable  \textit{OTT revenues}, Table \ref{OTT revenues} illustrates the time series of \textit{OTT revenues} in millions of US dollars in the considered countries, over the reference period 2008-2013, obtained by summing the \textit{OTT revenues} determined for every single OTT company.

\begin{center}
\renewcommand{\arraystretch}{1.1}
\begin{longtable}[h!]{K{0.28\textwidth}|K{0.28\textwidth}|K{0.28\textwidth}}
\caption{OTT revenues (in millions of US dollars)}\\
\label{OTT revenues}
\textbf{Year} & \textbf{Country} & \textbf{Annual revenues}  \\
\hline \hline\\[-2.8ex]\hline
2008 & Japan  & 0.05 \\ 
2009 & Japan       & 2.30 \\ 
2010 & Japan       & 19.87 \\ 
2011 & Japan       & 60.19 \\ 
2012 & Japan       & 113.55 \\ 
2013 & Japan       & 165.20 \\ 
2008 & USA   & 207.84 \\ 
2009 & USA  & 690.03 \\ 
2010 &  USA    & 1224.60 \\ 
2011 &  USA    & 2068.96 \\ 
2012 & USA            & 2580.53 \\ 
2013 & USA            & 3616.70 \\ 
2008 & UK          & 22.53 \\ 
2009 & UK          & 40.57 \\ 
2010 & UK          & 77.20 \\ 
2011 & UK          & 113.93 \\ 
2012 & UK          & 138.33 \\ 
2013 & UK          & 194.40 \\
2008 & France        & 12.99 \\ 
2009 & France        & 32.04 \\ 
2010 & France        & 72.44 \\ 
2011 & France        & 102.24 \\ 
2012 & France        & 125.21 \\ 
2013 & France        & 169.41 \\
2008 & Italy        & 9.24 \\ 
2009 & Italy        & 27.92 \\ 
2010 & Italy        & 59.83 \\ 
2011 & Italy        & 96.50 \\ 
2012 &  Italy        & 113.47 \\ 
2013 & Italy        & 149.94 \\
2008 & Spain        & 4.60 \\ 
2009 & Spain        & 25.35 \\ 
2010 & Spain        & 49.47 \\ 
2011 & Spain        & 71.27 \\ 
2012 & Spain        & 86.25 \\
2013 & Spain        & 117.74 \\ 
2008 & Germany        & 2.27 \\ 
2009 & Germany        & 20.80 \\ 
2010 & Germany        & 59.03 \\ 
2011 & Germany        & 97.54 \\ 
2012 &  Germany        & 123.50 \\
2013 &  Germany        & 162.41 \\
\end{longtable}
\end{center}

\section{Internet penetration}

With regard to the \textit{Internet Penetration}, firstly, the time series called ``\textit{Internet users (per 100 people)}" (last updated date 01.07.2015) has been obtained from \cite{InternetUsers}. According to the description of the data (available in \cite{InternetUsers}), the latter consists of the Internet users, i.e., the number of people by country with access to the worldwide network, per 100 people. Therefore, in order to obtain the time series of the \textit{Total Internet users} by country, the time series of the \textit{Total Population} in the considered countries has been employed. In particular, the time series called ``\textit{Total Population (in number of people)}"(last updated date 01.07.2015) has been acquired from \cite{TotalPopulation} and, in order to express the \textit{Internet users (per 100 people)} in terms of  \textit{Total Internet users}, the unknown of the following proportion has been determined:
\begin{eqnarray}\label{eq 2}
Internet \ users \ (per \ 100 \ people) : 100 = x : Total \ Population \ (in \ number \ of \ people),
\end{eqnarray}
where $x$ is the variable \textit{Total Internet users}.  Moreover, the series of \textit{Total Internet users} resulting from (\ref{eq 2}) has been divided by 1 million in order to make it consistent with all the other variables of the dataset, which are expressed in millions. Table \ref{Internet penetration} illustrates the time series of the \textit{Internet Penetration}, i.e., the \textit{Total Internet users} (in millions of people) in the considered countries, over the reference period 2008-2013.

\clearpage
\begin{center}
\renewcommand{\arraystretch}{1.1}
\begin{longtable}[h!]{K{0.28\textwidth}|K{0.28\textwidth}|K{0.28\textwidth}}
\caption{Internet penetration (in millions of people)}\\
\label{Internet penetration}
\textbf{Year} & \textbf{Country} & \textbf{Internet penetration} \\
\hline \hline\\[-2.8ex]\hline
2008 & Japan    &  96.56 \\ 
2009 & Japan    &  99.88 \\ 
2010 & Japan    &  100.16 \\ 
2011 & Japan    &  101.04 \\ 
2012 & Japan    &  110.02 \\ 
2013 & Japan    &  109.83 \\ 
2008 & USA      &  225.03 \\ 
2009 & USA      &  217.81 \\ 
2010 & USA      &  221.77 \\ 
2011 & USA      &  217.36 \\ 
2012 & USA      & 249.09 \\ 
2013 & USA      &  266.49 \\ 
2008 & UK       &  48.45 \\ 
2009 & UK       &  52.04 \\ 
2010 & UK       &  53.35 \\ 
2011 & UK       &  54.01 \\ 
2012 & UK       &  55.73 \\ 
2013 & UK       &  57.60 \\
2008 & France   &  45.50 \\ 
2009 & France   &  46.31 \\ 
2010 & France   &  50.25 \\ 
2011 & France   &  50.85 \\ 
2012 & France   &  53.45 \\ 
2013 & France   &  54.00 \\
2008 & Italy    &  26.20 \\ 
2009 & Italy    &  28.86 \\ 
2010 & Italy    &  31.82 \\ 
2011 & Italy    &  32.30 \\ 
2012 & Italy    &  33.24 \\ 
2013 & Italy    &  35.21 \\
2008 & Spain    &  27.39 \\ 
2009 & Spain    &  28.93 \\ 
2010 & Spain    &  30.65 \\ 
2011 & Spain    &  31.60 \\ 
2012 & Spain    &  32.65 \\ 
2013 & Spain    &  33.37 \\
2008 & Germany  &  64.05 \\ 
2009 & Germany  &  64.70 \\ 
2010 & Germany  &  67.06 \\ 
2011 & Germany  &  66.48 \\ 
2012 & Germany  &  66.23 \\ 
2013 & Germany  &  67.71 \\
\end{longtable}
\end{center}

\section{GDP}

Regarding the \textit{Gross Domestic Product (GDP)}, firstly, the time series called ``GDP (current \$)" (last updated date 28.07.2015) has been obtained from \cite{GDPcurrentUSD}. According to the description of the data (available in \cite{GDPcurrentUSD}), the latter is expressed in nominal terms, since the GDP is expressed in current US dollars. Therefore, the downloaded variable has been made real by employing the \textit{Consumer Price Index ($CPI$)}. In particular, the time series called ``\textit{Consumer price index (2010 = 100)}"(last updated date 29.07.2015) has been acquired from \cite{CPI} and, in order to express the nominal $GDP$ in real terms, the time series of nominal GDP has been divided by the time series of $CPI$:

\begin{eqnarray}\label{eq 1}
real \ GDP = \frac{nominal \ GDP}{CPI}.
\end{eqnarray}

Moreover, the series of real $GDP$ resulting from (\ref{eq 1}) has been divided by 1 million in order to make it consistent with all the other variables of the dataset, which are expressed in millions. Table \ref{Real GDP} illustrates the time series of real $GDP$ in millions of US dollars in the considered countries, over the reference period 2008-2013.

\begin{center}
\renewcommand{\arraystretch}{1.1}
\begin{longtable}[h!]{K{0.28\textwidth}|K{0.28\textwidth}|K{0.28\textwidth}}
\caption{Real GDP (in millions of US dollars)}\\
\label{Real GDP}
\textbf{Year} & \textbf{Country} & \textbf{Real GDP} \\
\hline \hline \\[-2.8ex]\hline
2008 & Japan &  47494.46 \\ 
2009 & Japan &  49986.51 \\ 
2010 & Japan &  54953.86 \\ 
2011 & Japan &  59222.15 \\ 
2012 & Japan &  59735.92 \\ 
2013 & Japan &  49175.96 \\ 
2008 & USA   &  149064.03 \\ 
2009 & USA   &  146546.79 \\ 
2010 & USA   &  149643.72 \\ 
2011 & USA   &  150425.80 \\ 
2012 & USA   &  153510.86 \\ 
2013 & USA   &  156960.15 \\ 
2008 & UK    &  29457.44 \\ 
2009 & UK    &  23848.33 \\ 
2010 & UK    &  24078.57 \\ 
2011 & UK    &  24808.73 \\ 
2012 & UK    &  24340.93 \\ 
2013 & UK    &  24307.26 \\
2008 & France &  29707.00 \\ 
2009 & France &  27351.28 \\ 
2010 & France &  26469.95 \\ 
2011 & France &  28030.77 \\ 
2012 & France &  25755.61 \\ 
2013 & France &  26761.73 \\
2008 & Italy  &  24469.31 \\ 
2009 & Italy  &  22199.83 \\ 
2010 & Italy  &  21267.48 \\ 
2011 & Italy  &  22173.34 \\ 
2012 & Italy  &  19601.60 \\ 
2013 & Italy  &  19941.66 \\
2008 & Spain  &  16595.50 \\ 
2009 & Spain  &  15260.86 \\ 
2010 & Spain  &  14316.73 \\ 
2011 & Spain  &  14481.64 \\ 
2012 & Spain  &  12823.80 \\ 
2013 & Spain  &  12993.57 \\
2008 & Germany   &  38001.18 \\ 
2009 & Germany   &  34505.87 \\ 
2010 & Germany   &  34122.12 \\ 
2011 & Germany   &  36754.28 \\ 
2012 & Germany   &  33931.07 \\ 
2013 & Germany   &  35294.36 \\
\end{longtable}
\end{center}

\section{Conclusion}

In this report, we provided a summary for observational and estimated data regarding important variables (i.e., ISP revenues, CAPEX, OTT revenues, Internet penetration and GDP) that can be exploited for empirical econometric studies on the relationship between ISPs and OTT providers. Our data constitute a balanced panel of ten major ISPs and three popular OTT providers that provide their services in seven OECD countries for the period 2008-2013. In our future work, we plan to update and extend this dataset taking into account more companies (both ISPs and OTT providers) and countries.

\bibliographystyle{ieeetr}
\bibliography{references}

\begin{thebibliography}{10}

\bibitem{ma}
R.~T. Ma, J.~Lui, and V.~Misra, ``Evolution of the internet economic
  ecosystem,'' {\em IEEE/ACM Transactions on Networking (TON)}, vol.~23, no.~1,
  pp.~85--98, 2015.

\bibitem{coucheney}
P.~Coucheney, P.~Maill{\'e}, and B.~Tuffin, ``Impact of competition between
  isps on the net neutrality debate,'' {\em IEEE Transactions on Network and
  Service Management}, vol.~10, no.~4, pp.~425--433, 2013.

\bibitem{saucez}
D.~Saucez, S.~Secci, and C.~Barakat, ``On the incentives and incremental
  deployments of icn technologies for ott services,'' {\em IEEE Network},
  vol.~28, no.~3, pp.~20--25, 2014.

\bibitem{courcoubetis}
C.~Courcoubetis, K.~Sdrolias, and R.~Weber, ``Revenue models, price
  differentiation and network neutrality implications in the internet,'' {\em
  ACM SIGMETRICS Performance Evaluation Review}, vol.~41, no.~4, pp.~20--23,
  2014.

\bibitem{Exchange2015WorldBank}
{The World Bank}, ``{Monthly Monetary and Financial Statistics (MEI): Exchange
  rates (USD monthly averages)}.''
  \url{http://stats.oecd.org/index.aspx?DatasetCode=MEI_FIN}.
\newblock [Online; Data extracted on 13.02.2015].

\bibitem{NTTDoCoMo2015NTTDoCoMo}
{NTT DoCoMo}, ``{NTT DoCoMo annual report 2014}.''
  \url{https://www.nttdocomo.co.jp/english/corporate/ir/binary/pdf/library/annual/fy2013/docomo_ar2014_e.pdf},
  2015.
\newblock [Online].

\bibitem{Softbank2014Softbank}
{Softbank}, ``{Softbank annual report 2013}.''
  \url{http://cdn.softbank.jp/en/corp/set/data/irinfo/financials/annual_reports/pdf/2013/softbank_annual_report_2013_001.pdf},
  2014.
\newblock [Online].

\bibitem{Softbank2015Softbank}
{Softbank}, ``{Softbank annual report 2014}.''
  \url{http://cdn.softbank.jp/en/corp/set/data/irinfo/financials/annual_reports/pdf/2014/softbank_annual_report_2014_001.pdf},
  2015.
\newblock [Online].

\bibitem{ATandT2011ATandT}
{AT \& T}, ``{At \& T annual report 2010}.''
  \url{http://www.att.com/Common/about_us/annual_report/pdfs/ATT2010_Full.pdf},
  2011.
\newblock [Online].

\bibitem{ATandT2014ATandT}
{AT \& T}, ``{At \& T annual report 2013}.''
  \url{http://www.att.com/Investor/ATT_Annual/2013/downloads/ar2013_annual_report.pdf},
  2014.
\newblock [Online].

\bibitem{Verizon2011Verizon}
{Verizon}, ``{Verizon annual report 2010}.''
  \url{http://www.verizon.com/about/investors/annual-report}, 2011.
\newblock [Online].

\bibitem{Verizon2014Verizon}
{Verizon}, ``{Verizon annual report 2013}.''
  \url{http://www.verizon.com/about/investors/annual-report}, 2014.
\newblock [Online].

\bibitem{BTGroup2011BTGroup}
{BTGroup}, ``{Verizon annual report 2010}.''
  \url{https://www.btplc.com/Sharesandperformance/Annualreportandreview/pdf/BTGroupAnnualReport2010.pdf},
  2011.
\newblock [Online].

\bibitem{BTGroup2014BTGroup}
{BTGroup}, ``{Verizon annual report 2013}.''
  \url{http://www.btplc.com/Sharesandperformance/Annualreportandreview/pdf/2013_BT_Annual_Report_smart.pdf},
  2014.
\newblock [Online].

\bibitem{Vodafone2011Vodafone}
{Vodafone}, ``{Vodafone annual report 2010}.''
  \url{http://www.vodafone.com/content/dam/vodafone/investors/annual_reports/annual_report_accounts_2010.pdf},
  2011.
\newblock [Online].

\bibitem{Vodafone2014Vodafone}
{Vodafone}, ``{Vodafone annual report 2013}.''
  \url{http://www.vodafone.com/content/annualreport/annual_report13/downloads/vodafone_annual_report_2013.pdf},
  2014.
\newblock [Online].

\bibitem{Orange2009Orange}
{Orange}, ``{Orange annual report 2008}.''
  \url{http://www.orange.com/en/content/download/4570/65453/version/2/file/annual-report2008_en.pdf},
  2009.
\newblock [Online].

\bibitem{Orange2010Orange}
{Orange}, ``{Orange annual report 2009}.''
  \url{http://www.orange.com/en/content/download/4572/65461/version/2/file/FTEL_1005297_complet_GB.indd_RVB.pdf},
  2010.
\newblock [Online].

\bibitem{Orange2011Orange}
{Orange}, ``{Orange annual report 2010}.''
  \url{http://www.orange.com/en/content/download/4548/65268/version/3/file/2010annualreport.pdf},
  2011.
\newblock [Online].

\bibitem{Orange2012Orange}
{Orange}, ``{Orange annual report 2011}.''
  \url{http://www.orange.com/en/content/download/6366/93028/version/3/file/RA2011_EN.pdf},
  2012.
\newblock [Online].

\bibitem{Orange2013Orange}
{Orange}, ``{Orange annual report 2012}.''
  \url{http://www.orange.com/en/content/download/12910/269508/version/2/file/Orange-RA-2012-GB.pdf},
  2013.
\newblock [Online].

\bibitem{Orange2014Orange}
{Orange}, ``{Orange annual report 2013}.''
  \url{http://www.orange.com/en/content/download/23308/480043/version/2/file/the_little_Orange_book.pdf},
  2014.
\newblock [Online].

\bibitem{Telecom2010Telecom}
{Telecom Italia}, ``{Telecom Italia annual report 2009}.''
  \url{https://www.telecomitalia.com/content/dam/telecomitalia/en/archive/documents/investors/Annual_Reports2009/2009Annual_Report.pdf},
  2010.
\newblock [Online].

\bibitem{Telecom2011Telecom}
{Telecom Italia}, ``{Telecom Italia annual report 2010}.''
  \url{http://www.telecomitalia.com/content/dam/telecomitalia/en/archive/documents/investors/Annual_Reports/2010/AnnualReport2010.pdf},
  2011.
\newblock [Online].

\bibitem{Telecom2012Telecom}
{Telecom Italia}, ``{Telecom Italia annual report 2011}.''
  \url{http://www.telecomitalia.com/content/dam/telecomitalia/en/archive/documents/investors/Annual_Reports/2011/TelecomItaliaGroupAnnualReport2011.pdf},
  2012.
\newblock [Online].

\bibitem{Telecom2013Telecom}
{Telecom Italia}, ``{Telecom Italia annual report 2012}.''
  \url{http://www.telecomitalia.com/content/dam/telecomitalia/en/archive/documents/investors/Annual_Reports/2012/AnnualReport2012.pdf},
  2013.
\newblock [Online].

\bibitem{Telecom2014Telecom}
{Telecom Italia}, ``{Telecom Italia annual report 2013}.''
  \url{http://www.telecomitalia.com/content/dam/telecomitalia/en/archive/documents/investors/Annual_Reports/2013/Annual-Report-2013.pdf},
  2014.
\newblock [Online].

\bibitem{Telefonica2010Telefonica}
{Telef\'onica}, ``{Telef\'onica annual report 2009}.''
  \url{http://www.telefonica.com/en/about_telefonica/pdf/informes/2009/Telefonica_IA09_Ing.pdf},
  2010.
\newblock [Online].

\bibitem{Telefonica2011Telefonica}
{Telef\'onica}, ``{Telef\'onica annual report 2010}.''
  \url{http://www.telefonica.com/en/about_telefonica/pdf/informes/2010/telefonica_ia10_eng.pdf},
  2011.
\newblock [Online].

\bibitem{Telefonica2014Telefonica}
{Telef\'onica}, ``{Telef\'onica annual report 2013}.''
  \url{http://www.telefonica.com/en/shareholders-investors/pdf/20140320_Consolidated_Annual_Accounts_311213.pdf},
  2014.
\newblock [Online].

\bibitem{DeutscheTelecom2011DeutscheTelecom}
{Deutsche Telecom}, ``{Deutsche Telecom annual report 2010}.''
  \url{http://www.telekom.com/investor-relations/publications/Financial-results/205540},
  2011.
\newblock [Online].

\bibitem{DeutscheTelecom2014DeutscheTelecom}
{Deutsche Telecom}, ``{Deutsche Telecom annual report 2013}.''
  \url{http://www.telekom.com/ar-2013}, 2014.
\newblock [Online].

\bibitem{Fujimoto2013Marketing}
G.~Fujimoto, ``{Marketing I Skype HOMEpage}.''
  \url{http://www.slideshare.net/goc1126/skype-marketing-final-28623964}, 2013.
\newblock [Online; published on 25.11.2013].

\bibitem{Mercier2013Skype}
J.~Mercier, ``{Skype Numerology}.''
  \url{http://skypenumerology.blogspot.ch/2012/01/skype-calling-minutes-in-2011.html},
  2012.
\newblock [Online; published on 19.01.2012].

\bibitem{Paraskelidis2013Evaluating}
M.~A. Athanasios~Paraskelidis and M.~T. P.~M. Dewage, ``{Evaluating the Energy
  Efficiency of Modern VoIP Applications}.''
  \url{http://www.atiner.gr/papers/COM2013-0787.pdf}, 2013.
\newblock [Online; published on 20.12.2013].

\bibitem{Warren2014Viber}
T.~Warren, ``{Viber messaging app acquired by Japan's Rakuten for \$ 900
  million}.''
  \url{http://www.theverge.com/2014/2/14/5411082/viber-messaging-app-acquired-by-rakuten},
  2014.
\newblock [Online; published on 14.02.2014].

\bibitem{eMarketer2013Three}
{eMarketer}, ``{Three Out of Four UK Mobile Users to Own Smartphones by
  2016}.''
  \url{http://www.emarketer.com/Article/Three-of-Four-UK-Mobile-Users-Own-Smartphones-by-2016/1009614},
  2013.
\newblock [Online; published on 18.01.2013].

\bibitem{furrer2016corporate}
O.~Furrer, {\em Corporate level strategy: Theory and applications}.
\newblock Routledge, 2016.

\bibitem{Statista2016Skype}
Statista, ``{Skype's annual revenue from 2006 to 2010 (in million U.S.
  dollars)}.''
  \url{https://www.statista.com/statistics/266191/skype-revenue-since-2006/},
  2016.

\bibitem{Bass2013Microsoft}
D.~Bass, ``{Microsoft Skype Unit Approaching \$2 Billion in Annual Sales}.''
  \url{https://www.bloomberg.com/news/articles/2013-02-19/microsoft-s-skype-unit-approaching-2-billion-in-annual-revenue},
  2013.
\newblock [Online; published on 19.02.2013].

\bibitem{Sedghi2014Facebook}
A.~Sedghi, ``{Facebook: 10 years of social networking, in numbers}.''
  \url{https://www.theguardian.com/news/datablog/2014/feb/04/facebook-in-numbers-statistics},
  2014.
\newblock [Online; published on 04.02.2014].

\bibitem{Tabuchi2011Facebook}
H.~Tabuchi, ``{Facebook Wins Relatively Few Friends in Japan}.''
  \url{http://www.nytimes.com/2011/01/10/technology/10facebook.html}, 2011.
\newblock [Online; published on 09.01.2011].

\bibitem{hamada2012facebook}
M.~Hamada, ``A facebook project for japanese university students: Does it
  really enhance student interaction, learner autonomy, and english
  abilities?,'' in {\em EUROCALL Conference}, p.~104, 2012.

\bibitem{Darwell2012Facebook}
B.~Darwell, ``{Facebook reaches 10M in Japan, doubles users in 6 months}.''
  \url{http://www.adweek.com/socialtimes/facebook-reaches-10m-in-japan-doubled-users-in-6-months/276511},
  2012.
\newblock [Online; published on 16.03.2012].

\bibitem{Bloomberg2012Japan}
{Bloomberg}, ``{In Japan Facebook wins the most users}.''
  \url{http://www.bloomberg.com/news/articles/2012-03-22/in-japan-facebook-wins-the-most-users},
  2012.
\newblock [Online; published on 22.03.2012].

\bibitem{Statista2014Number}
{Statista}, ``{Number of Facebook users in Japan}.''
  \url{https://www.statista.com/statistics/304831/number-of-facebook-users-in-japan/},
  2014.

\bibitem{SocialTimes2008Latest}
{SocialTimes}, ``{Latest Data on US Facebook Age and Gender Demographics}.''
  \url{http://www.adweek.com/socialtimes/latest-data-on-us-facebook-age-and-gender-demographics/213798},
  2008.
\newblock [Online; published on 18.09.2008].

\bibitem{SocialTimes2010December}
{SocialTimes}, ``{December Data on Facebook's US Growth by Age and Gender:
  Beyond 100 Million}.''
  \url{http://www.adweek.com/socialtimes/december-data-on-facebook%E2%80%99s-us-growth-by-age-and-gender-beyond-100-million/233478},
  2010.
\newblock [Online; published on 04.01.2010].

\bibitem{Eldon2010Facebook}
E.~Eldon, ``{Facebook US Demographic Data for September 2010: Nearly 5 Million
  More Users}.''
  \url{http://www.adweek.com/socialtimes/facebook-us-demographic-data-for-september-2010-nearly-5-million-more-users/248737},
  2010.
\newblock [Online; published on 04.10.2010].

\bibitem{Lee2011Facebook}
A.~Lee, ``{Facebook Users DROP In U.S.: Millions Left The Social Network In May
  2011}.''
  \url{http://www.huffingtonpost.com/2011/06/13/facebook-users-members-us-growth-drops-may-2011_n_875810.html},
  2011.
\newblock [Online; published on 13.06.2011].

\bibitem{Kiss2013Facebook}
J.~Kiss, ``{Facebook UK loses 600,000 users in December}.''
  \url{https://www.theguardian.com/technology/2013/jan/14/facebook-loses-uk-users-december},
  2013.
\newblock [Online; published on 14.01.2013].

\bibitem{Saul2014MillionTeens}
D.~Saul, ``{3 Million Teens Leave Facebook In 3 Years: The 2014 Facebook
  Demographic Report}.''
  \url{https://isl.co/2014/01/3-million-teens-leave-facebook-in-3-years-the-2014-facebook-demographic-report/},
  2014.
\newblock [Online; published on 15.01.2014].

\bibitem{WikidotSocial}
{Wikidot}, ``{Social Media Statistics}.''
  \url{http://socialmediastatistics.wikidot.com/facebook}.
\newblock [Online].

\bibitem{Burcher2010Facebook}
N.~Burcher, ``{Facebook usage statistics by country - July 2010 compared to
  July 2009 and July 2008}.''
  \url{http://www.nickburcher.com/2010/07/facebook-usage-statistics-by-country.html},
  2010.
\newblock [Online; published on 02.07.2010].

\bibitem{eMarketer2013Emerging}
{eMarketer}, ``{Emerging Markets Drive Facebook User Growth}.''
  \url{https://www.emarketer.com/Article/Emerging-Markets-Drive-Facebook-User-Growth/1009875},
  2013.
\newblock [Online; published on 09.05.2013].

\bibitem{SocialTimes2009Bonjour}
{SocialTimes}, ``{Bonjour! Inside Facebook France Launches Today en
  Français}.''
  \url{http://www.adweek.com/socialtimes/bonjour-inside-facebook-france-launches-today-en-francais/216052},
  2009.
\newblock [Online; published on 04.01.2009].

\bibitem{Rao2011Facebook}
L.~Rao, ``{Facebook Now Has 149M Active Users In The U.S.; 70 Percent Log On
  Daily}.''
  \url{https://techcrunch.com/2011/02/10/facebook-now-has-149m-active-users-in-the-u-s-70-percent-log-on-daily/},
  2011.
\newblock [Online; published on 10.02.2011].

\bibitem{Yeates2012UK}
O.~Yeates, ``{UK facebook statistics february 2012}.''
  \url{https://www.clicky.co.uk/2012/02/uk-facebook-statistics-february-2012/},
  2012.
\newblock [Online; published on 28.02.2012].

\bibitem{InternetInternet}
{The Internet Coaching Library: Telecommunications Research Reports},
  ``{Internet World Stats: Usage and Population Statistics}.''
  \url{http://www.internetworldstats.com/europa.htm}.
\newblock [Online].

\bibitem{Constine2013Facebook}
J.~Constine, ``{Facebook's Cutesy Annual Report To Partners Reveals First
  Country-By-Country Mobile Stats}.''
  \url{https://techcrunch.com/2013/12/29/facebook-international-user-growth/},
  2013.
\newblock [Online; published on 29.12.2013].

\bibitem{Mysore2008FacebookItaly}
S.~Mysore, ``{Facebook Growth Surges in Italy, Developers Look for Better
  Italian eCPMs}.''
  \url{http://www.adweek.com/socialtimes/facebook-growth-surges-in-italy-developers-look-for-better-italian-ecpms/215697},
  2008.
\newblock [Online; published on 18.12.2008].

\bibitem{YungHui2012Billion}
L.~Yung-Hui, ``{1 Billion Facebook Users On Earth: Are We There Yet?}.''
  \url{http://www.forbes.com/sites/limyunghui/2012/09/30/1-billion-facebook-users-on-earth-are-we-there-yet/\string#751363872a0e},
  2012.
\newblock [Online; published on 30.09.2012].

\bibitem{Mysore2008Footprint}
S.~Mysore, ``{Facebook's Footprint in Spain Up 600\% in 2008}.''
  \url{http://www.adweek.com/socialtimes/facebooks-footprint-in-spain-up-600-in-2008/215705},
  2008.
\newblock [Online; published on 19.12.2008].

\bibitem{Parkinson2011Tuenti}
L.~Parkinson, ``{How Tuenti Held Off Facebook in Spain with Better Privacy}.''
  \url{http://mediashift.org/2011/03/how-tuenti-held-off-facebook-in-spain-with-better-privacy068/},
  2011.
\newblock [Online; published on 09.03.2011].

\bibitem{LATEVAWEBHowMany}
LATEVAWEB, ``{How many people use Facebook in Spain}.''
  \url{https://www.latevaweb.com/usodefacebook-en.html}.
\newblock [Online].

\bibitem{Mysore2008Germany}
S.~Mysore, ``{Facebook Germany Reaches More Than 1M Users, Gaining Ground on
  Competitor StudiVZ}.''
  \url{http://www.adweek.com/socialtimes/facebook-germany-reaches-more-than-1m-users-gaining-ground-on-competitor-studivz/215540},
  2008.
\newblock [Online; published on 18.12.2008].

\bibitem{Socialnomics2013Facebook}
{Socialnomics}, ``{Facebook and Germany to like or not to like}.''
  \url{http://socialnomics.net/2013/04/24/facebook-and-germany-to-like-or-not-to-like/},
  2013.
\newblock [Online; published on 24.04.2013].

\bibitem{Facebook2013Facebook}
{Facebook, Inc.}, ``{Facebook annual report 2012}.''
  \url{https://materials.proxyvote.com/Approved/30303M/20130409/AR_166822/document.pdf},
  2013.
\newblock [Online].

\bibitem{Facebook2014Facebook}
{Facebook, Inc.}, ``{Facebook annual report 2013}.''
  \url{https://materials.proxyvote.com/Approved/30303M/20140324/AR_200747/pubData/source/Facebook%20AR%204-1-14.pdf},
  2014.
\newblock [Online].

\bibitem{Flore2015Things}
E.~Flore, ``{5 Things You Can Learn From The Story Of WhatsApp}.''
  \url{https://medium.com/the-story-of-grip/5-things-every-founder-can-learn-from-the-story-of-whatsapp-b6496bc4f54d\string#.lk67xjuku},
  2015.
\newblock [Online; published on 29.06.2015].

\bibitem{Satariano2014WhatsApp}
A.~Satariano, ``{WhatsApp's Founder Goes From Food Stamps to Billionaire}.''
  \url{https://www.bloomberg.com/news/articles/2014-02-20/whatsapp-s-founder-goes-from-food-stamps-to-billionaire},
  2014.
\newblock [Online; published on 20.02.2014].

\bibitem{Bradshaw2011WhatsApp}
T.~Bradshaw, ``{WhatsApp users get the message}.''
  \url{https://www.ft.com/content/30fd99a2-0c60-11e1-88c6-00144feabdc0\string#axzz3SqjQS5NT},
  2011.
\newblock [Online; published on 14.11.2011].

\bibitem{Lomas2013Skype}
N.~Lomas, ``{Skype Competitor Viber Hits 175 Million Users, Up From 140
  Million+ In December}.''
  \url{https://techcrunch.com/2013/02/26/skype-competitor-viber-hits-175-million-users-up-from-140-million-in-december/},
  2013.
\newblock [Online; published on 26.02.2013].

\bibitem{Rowan2014WhatsApp}
D.~Rowan, ``{WhatsApp: The inside story}.''
  \url{http://www.wired.co.uk/article/whatsapp-exclusive}, 2014.
\newblock [Online; published on 19.02.2014].

\bibitem{Statista2014Annual}
Statista, ``{Annual revenue of WhatsApp from 2012 to 1st half 2014 (in million
  U.S. dollars)}.''
  \url{https://www.statista.com/statistics/346269/whatsapp-annual-revenue/},
  2014.
\newblock [Online].

\bibitem{InternetUsers}
{The World Bank}, ``{World Development Indicators: Internet users (per 100
  people)}.'' \url{http://data.worldbank.org/indicator/IT.NET.USER.P2}.
\newblock [last updated date: 01.07.2015].

\bibitem{TotalPopulation}
{The World Bank}, ``{World Development Indicators: Total Population (in number
  of people)}.'' \url{http://data.worldbank.org/indicator/SP.POP.TOTL}.
\newblock [last updated date: 01.07.2015].

\bibitem{GDPcurrentUSD}
{The World Bank}, ``{World Development Indicators: GDP (current USD)}.''
  \url{http://data.worldbank.org/indicator/NY.GDP.MKTP.CD}.
\newblock [last updated date: 28.07.2015].

\bibitem{CPI}
{The World Bank}, ``{World Development Indicators: Consumer price index (2010 =
  100)}.'' \url{http://data.worldbank.org/indicator/FP.CPI.TOTL}.
\newblock [last updated date: 28.07.2015].

\end{thebibliography}

\end{document}